\documentclass[12pt]{article}
\usepackage{fancyhdr}

\usepackage{overcite}

\usepackage{graphicx}

\usepackage{amsfonts}

\usepackage{amsmath}

\usepackage{siunitx}

\usepackage{epstopdf}

\usepackage{caption}

\usepackage{subcaption}

\usepackage{authblk}

\usepackage{setspace}

\usepackage{multirow}

\usepackage{tabularx}

\usepackage{makecell}

\usepackage[utf8]{inputenc}

\usepackage[mathlines]{lineno}
\usepackage{rotating}

\usepackage{soul}
\usepackage{xcolor}
\makeatletter \renewcommand\@biblabel[1]{$^{#1}$} \makeatother
\newlength{\bibhang}
\title{Technical Note: MC-GPU breast dosimetry validations with other Monte Carlo codes and Phase Space File implementation}
\author{Rodrigo T. Massera}
\author{Rowan M. Thomson}
\author{Alessandra Tomal}
\newcommand{\DgNCT}{DgN$_{\text{CT}}$}
\newcommand{\Kair}{K$_{\text{air}}$}

\newcommand{\cen}[1]{\begin{center} #1 \end{center}}

\begin{document}

\cen{\sf {\Large {\bfseries Technical Note: MC-GPU breast dosimetry validations with other Monte Carlo codes and Phase Space File implementation} \\
Rodrigo T. Massera$^\text{1,2}$, Rowan M. Thomson$^\text{2}$, Alessandra Tomal$^\text{1}$} \\ \vspace{3mm}
1: Instituto de Física "Gleb Wataghin", Universidade Estadual de Campinas, Campinas, São Paulo, 13083-859,  Brazil\\
2: Carleton Laboratory for Radiotherapy Physics, Department of Physics, Carleton University, Ottawa, Ontario, K1S 5B6, Canada \\ \vspace{5mm} 
Author to whom correspondence should be addressed. email: atomal$@$ifi.unicamp.br \\ \vspace{5mm}
%submitted on March 8, 2020 \\
}
\pagestyle{empty}

%Efficiency at 1 MeV
% Single cell
%    1/3.01727e+07/(25.328)^2=5.1663524e-11
% Recursive cells
%    1/2.15654e+06/(0.027)^2=0.0006360847
%
% Ratio
%    12312065.6655 (higher than 12 million)

%           _         _                  _   
% A limit of 300 words applies to Technical Notes and Medical Physics Letters.
\begin{abstract}
\noindent\textbf{Purpose:\ }To validate the MC-GPU Monte Carlo code for dosimetric studies in  x-ray breast imaging modalities:~mammography, digital breast tomosynthesis, contrast enhanced digital mammography and breast-CT. Moreover, to implement and validate a phase space file generation routine.

\noindent\textbf{Methods:\ }The MC-GPU code (v.~1.5 DBT) was modified in order to generate phase space files and to be compatible with PENELOPE v.~2018 derived cross section database. Simulations were performed with homogeneous and anthropomorphic breast phantoms
for different breast imaging techniques. The glandular dose was computed for each case and  compared with results from the PENELOPE (v.~2014) + penEasy (v.~2015) and egs$\_$brachy (EGSnrc) Monte Carlo codes. Afterwards, several phase space files were generated with MC-GPU and the scored photon spectra were compared between the codes. The phase space files generated in MC-GPU were used in PENELOPE and EGSnrc to calculate the glandular dose, and compared with the original dose scored in MC-GPU.
  
\noindent\textbf{Results:\ }MC-GPU showed good agreement with the other codes when calculating the glandular dose distribution for mammography, mean glandular dose for digital breast tomosynthesis, and normalized glandular dose for breast-CT. The latter case showed average/maximum relative differences of 2.3\%/27\%, respectively, compared to other literature works, with the larger differences observed at low energies (around 10 keV). The recorded photon spectra entering a voxel were similar (within statistical uncertainties) between the three Monte Carlo codes. Finally, the reconstructed glandular dose in a voxel from a phase space file differs by less than 0.65\%, with an average of 0.18\% to 0.22\%  between the different MC codes, agreement within approximately $2\sigma$ statistical uncertainties. In some scenarios, the simulations performed in MC-GPU were from 20 up to 40 times faster than those performed by PENELOPE.

\noindent\textbf{Conclusions:\ }The results indicate that MC-GPU code is suitable for breast dosimetric studies  for different x-ray breast imaging modalities, with the advantage of a high performance derived from GPUs. The phase space file implementation was validated and is compatible with the IAEA standard, allowing multiscale Monte Carlo simulations with a combination of CPU and GPU codes.

\end{abstract}

\newpage
% Good for reviewing
\setlength{\baselineskip}{0.7cm}
\pagestyle{fancy}

%-------------------------------------- INTRODUCTION ----------------------
\section{Introduction} \label{sec:Introduction} 

Monte Carlo (MC) simulations are a powerful  tool employed for glandular dose assessments for x-ray breast imaging\cite{Dance1990,Wu1991,Boone1999,Sarno2018,Dance2016}. With advances in computational power and promising imaging techniques for studying breast anatomy, there is a growing interest in performing advanced dose evaluations in mammography and other related x-ray imaging techniques \cite{Mettivier2020, Hernandez2015, Hernandez2019}, such as mean glandular dose calculations and 3D dose distribution in anthropomorphic breast phantoms. The increase in complexity, mainly from  the realistic breast models, and a high  number of simulations (from hundreds to thousands possible combinations between parameters and models) requires considerable hardware resources and computational power. One option is to take advantage of central processing unit (CPU) parallelism capabilities and distribute the necessary MC calculations over a large number of CPUs. Another option, depending of the application, is to implement Graphical Processing Units (GPUs) to perform the calculations instead the traditional CPUs. With this approach, a single GPU could match the performance of several CPUs, as previously shown with the MC-GPU code \cite{Badal2009}, thus allowing complex simulations with reduced hardware resources. However, this MC code only simulates photon transport\cite{Badal2009}. MC-GPU has already been employed for simulating some applications involving low-energy (x-ray) beams, e.g.~breast imaging studies and virtual clinical trials\cite{Badano2018, Badal2020} and coherent x-ray scattering \cite{Ghammraoui2014} by adapting the code to include molecular interference \cite{Ghammraoui2014b}. Original and modified MC-GPU codes were also validated for applications in interventional radiology and cardiology \cite{FernandezBosman2021,Garcia2021}. 
MC-GPU was also adapted for patient specific CT dose calculations \cite{Sharma2019b}. In addition, traditional CPU Monte Carlo codes were adapted to GPU, e.g.~Geant4 \cite{Bert2013}  and EGSnrc \cite{Lippuner2011}. A GPU Monte Carlo code was also developed for DNA damage simulations due to ionizing radiation \cite{Tsai2020, Lai2020}. These examples demonstrate the capabilities of GPU MC codes and their possible applications. Nevertheless, to our knowledge, a framework for multiscale dose calculations in mammography x-ray imaging using a combination of GPU and CPU MC codes was not yet implemented.

MC-GPU has been used for breast imaging studies with a focus on image quality (due to its performance advantages), however, there has not been a detailed comparison between MC-GPU and other MC codes with a focus in breast dosimetry.  This would be useful, especially with the current developments in anthropomorphic phantoms for breast dosimetry, and could support migration from CPU MC codes to GPU ones.  
With recent interest in other x-ray breast imaging modalities besides mammography, an efficient and validated MC code capable of performing dosimetry studies in different modalities would be of interest.

With MC-GPU, the simulations are limited to macroscopic scales, where the geometric components ({\it e.g.}, voxels) are several times the electron range, and only photon transport is modeled.  This approximation is acceptable, for example, to estimate the mean glandular dose and 3D dose distributions ({e.g.}, in mm-scale voxels) \cite{Sechopoulos2015, Sarno2017b}. On the other hand, a more detailed approach for dosimetric analysis in x-ray breast imaging involves multiple length scales, including cell populations \cite{Oliver2019}, for which electron transport must be considered.

One possible approach for these multiscale simulations is to segment the simulation into different steps. First, the GPU code could simulate photon transport in the macroscopic geometry ({\it e.g.}~in a virtual patient model) and then record the phase space information for particles entering a smaller region.  Next, a CPU code could be used to simulate coupled electron-photon transport within the smaller volume with more detailed microscopic model.
With this concept, MC-GPU could be employed in a multiscale framework for x-ray breast imaging dosimetry.

The present work focuses on developments that are relevant for application of MC-GPU for breast dosimetry and is divided in two main parts:~the first one describes a detailed validation with MC-GPU for breast dosimetry considering different imaging modalities:\  mammography, digital breast tomosynthesis (DBT) and breast-CT. Meanwhile, the second part consists of an implementation and validation of the phase space file generation algorithm which includes the previous cited imaging modalities plus contrast-enhanced digital mammography (CEDM).

%--------------------------------------

%-------------------------------------- METHODS ----------------------

\section{Methods} \label{sec:Methods}

The MC-GPU (v.\ 1.5 VICTRE-DBT) \cite{Badal2021} code was employed with some modifications. This code uses the interaction scoring method.  The cross section database was updated from PENELOPE 2006 to the newer version 2018 \cite{Salvat2019}. For comparison purposes, two other codes were used:\ the previously modified and validated\cite{Massera2018} PENELOPE\cite{Salvat2019} (v.~2014) + penEasy\cite{Sempau2011} (v.~2015); and egs$\_$brachy \cite{Chamberland2016}, an application of EGSnrc. For PENELOPE 2014, the default cross section database was implemented (which is similar to the 2018 version)  and the interaction scoring was used, while for EGSnrc the mcdf-XCOM photon cross section with the PENELOPE energy absorption coefficients were used (calculated using PENELOPE routines) with tracklength scoring. The statistical uncertainties were estimated using the history-by-history method, which updates the uncertainty counters at the end of each primary particle history (more details in Refs. \cite{Sempau2001, Walters2002}).

Electron transport was not modeled.  The photon energy cutoff was set to 1 keV.
Information regarding the material compositions and the respective references are contained in Table \ref{tab:compos}.
%The glandular, adipose and skin breast tissues composition were extracted from Hammerstein et al.\ \cite{Hammerstein1979}, while the connective tissue was taken from Woodard and White \cite{Woodard1986}. The remaining material compositions were retrieved from NIST \cite{Hubbell2004}. More information regarding the material compositions are contained in %appendix \ref{ap:elemental_compositions}
%Table \ref{tab:compos}.
\vspace{10mm}

\begin{table}[htpb]
\caption{Elemental composition (in mass percent composition) of the materials employed in the simulations with their respective reference.}
\label{tab:compos}
\begin{center}
\begin{small}
\begin{tabular}{ c c c c c c c }
\hline
\hline
 Material & \multicolumn{1}{p{1.5cm}}{Density (g/cm$^{3}$)} & H & C & N & O & \multicolumn{1}{p{6.0cm}}{Others} \\ 
 \hline
 Adipose\cite{Hammerstein1979} & 0.93 & 11.2 & 61.9 & 1.7 & 25.1 &\multicolumn{1}{p{5.0cm}}{P(0.025),S(0.025),K(0.025),Ca(0.025)}\\
 Glandular\cite{Hammerstein1979} & 1.04 & 10.2 & 18.4 & 3.2 & 67.7 &\multicolumn{1}{p{5.0cm}}{P(0.125),S(0.125),K(0.125),Ca(0.125)}\\
 Skin\cite{Hammerstein1979} & 1.09 & 9.8 & 17.8 & 5.0 & 66.7 &\multicolumn{1}{p{5.0cm}}{P(0.175),S(0.175),K(0.175),Ca(0.175)}\\
  Connective\cite{Woodard1986} &  1.12 & 9.4 & 20.7 & 6.2 & 62.2 &\multicolumn{1}{p{5.0cm}}{Na(0.2),S(0.6),Cl(0.3)}\\
  Blood (ICRP) \cite{Hubbell2004} & 1.06 & 10.187 & 10.002 & 2.964 & 75.941 &\multicolumn{1}{p{5.0cm}}{Na(0.185),Mg(0.004),Si(0.003), P(0.035),S(0.185),Cl(0.278),K(0.163), Ca(0.006),Fe(0.046),Zn(0.001)}\\
 Muscle (ICRP) \cite{Hubbell2004}& 1.04 & 10.064 & 10.783 & 2.768 & 75.477 &\multicolumn{1}{p{5.0cm}}{Na(0.075),Mg(0.019),P(0.180),S(0.241), Cl(0.079),K(0.302),Ca(0.003), Fe(0.004),Zn(0.005)}\\
 PMMA\cite{Hubbell2004} & 1.19 & 8.054 & 59.985 & - & 31.961 & - \\
 \hline
 \hline
\end{tabular}
\end{small}
\end{center}
\end{table}

The simulations using PENELOPE were performed in a Ryzen 1700x (AMD, USA) and Core i7 7700 (Intel, USA), while for MC-GPU the simulations were performed in a GeForce GTX 1060 (NVIDIA, USA).

\subsection{Dosimetry validations}

This section covers the dosimetry validation for  different breast imaging modalities and breast models. Subsection \ref{sec:mgd_dgn_val} includes the digital breast tomosynthesis (DBT) and breast-CT validations for homogeneous breast models, while subsection \ref{sec:method_mgd} describes the validation for mammography using heterogeneous breast models. Table \ref{tab:tomo_bct_mammo_val} summarizes the general parameters employed in the simulations explained further.
\vspace{5mm}

\begin{table}[htpb]
\caption{Overview of simulations for dosimetric validations:\ breast geometric descriptors (shape, radius, thickness) and glandularity, source parameters (x-ray spectra, field size, source-detector/isocenter distances), and scored quantities for each simulated modality, as well as the publication motivating the simulation.}
\label{tab:tomo_bct_mammo_val}
\begin{center}
\begin{small}
\begin{tabular}{lccc}
\hline
\hline
                     & \multicolumn{3}{c}{Simulated modality}             \\
                     \cline{2-4}
Parameter  & DBT & Breast-CT & Mammography        \\
\hline
Breast shape              & Semicylinder                 & Cylinder  &  Semicylinder     \\
Breast radius            & 8 cm                         & 4, 6, 9 cm  & $\approx$ 10 cm       \\
Breast thickness (height)  & 2, 5, 8 cm                   & 4, 9, 18 cm & 5 cm     \\
Glandularity              & 1, 50, 100\%                 & 0.1, 50, 100\% & 20\%  \\
Field size               & 26 $\times$ 14 cm$^{2}$        & 40 $\times$ 30 cm$^{2}$ & 26 $\times$ 14 cm$^{2}$      \\
Source detector distance   & 66 cm                        & 92.3 cm &  66 cm       \\
Source isocenter distance   & 66 cm                        & 65 cm   & -         \\
X-ray spectra             & W/Rh:\ 23, 28, 35 kV           & Mono:\ 10 – 80 keV & W/Rh:\ 28 kV \\
Scored quantity           & MGD                          & DgN$_{CT}$   & DD*  \\
Adapted geometry from     & TG-195\cite{Sechopoulos2015}, TG-223\cite{Sechopoulos2014}  & Sarno et al.\cite{Sarno2018}  & TG-195\cite{Sechopoulos2015}  \\
\hline
\hline
\end{tabular}
\end{small}
\end{center}
\begin{footnotesize}
*DD: dose distribution.
\end{footnotesize}\vspace{5mm}
\end{table}

\subsubsection{Dosimetry validations for homogeneous breast models}

\label{sec:mgd_dgn_val}

 The DBT dosimetry validation consisted of two steps. First, a modified version of  PENELOPE/penEasy  MC code for breast dosimetry was validated against the report of Task Group 223\cite{Sechopoulos2014} (results of this step are available in the Supplementary Materials). Second, we compared the modified PENELOPE code  results with MC-GPU using a geometry adapted from Task Group 223.  The  MC-GPU geometry consisted of voxelized rectilinear geometries, which are the only geometries that may be simulated within this code, with 0.5 mm resolution. The adapted geometry is described as a 66 cm source-to-detector distance, and a 26~$\times$~14~cm$^{2}$ x-ray field at the detector entrance (Table \ref{tab:tomo_bct_mammo_val}). The support and compression plates (2~mm thick, PMMA) were also included. In MC-GPU and PENELOPE, the breast was modeled as a randomly sampled adipose-glandular distribution. 
 For both codes, the inner breast is surrounded by a 1.5~mm skin thickness, and the breast has a semi-cylindrical shape (8~cm radius) to address a cracioncaudal (CC) view. The skin is absent in  the region where the breast would be in contact with the chest wall. Three breast thickness/glandularity combinations were evaluated:\ 2~cm/100\%; 5~cm/50\%; 8 cm/1\%, whose selection was based on the extreme values usually employed in dose validation studies\cite{Boone1999, Sechopoulos2014}. The following spectra, obtained from TASMICS \cite{Hernandez2017}, were used: W/Rh 23~kV; W/Rh~28 kV; W/Rh~35 kV  for 2~cm, 5~cm and 8~cm breast thicknesses, respectively. 
 The detector-center of rotation distance was set to 0~cm, and the mean glandular dose (MGD,\ i.e.\ the sum of the  energy deposited in glandular voxels by their total mass) was compared between MC-GPU and PENELOPE codes from a 0$^{\circ}$ to 30$^{\circ}$ tube rotation angle (5$^{\circ}$ step). The 0$^{\circ}$ DBT projection presents a similar acquisition geometry of a mammography examination, thus a specific mammography validation for the homogeneous model was not included. The total number of  primary photons were in the order of 10$^{8}$ for PENELOPE and 10$^{9}$ for MC-GPU. We validated MC-GPU with PENELOPE and not with TG223 directly due to the difficulty to convert the geometry to voxels.

\begin{figure}[htpb]
\centering
\includegraphics[width=0.85\textwidth]{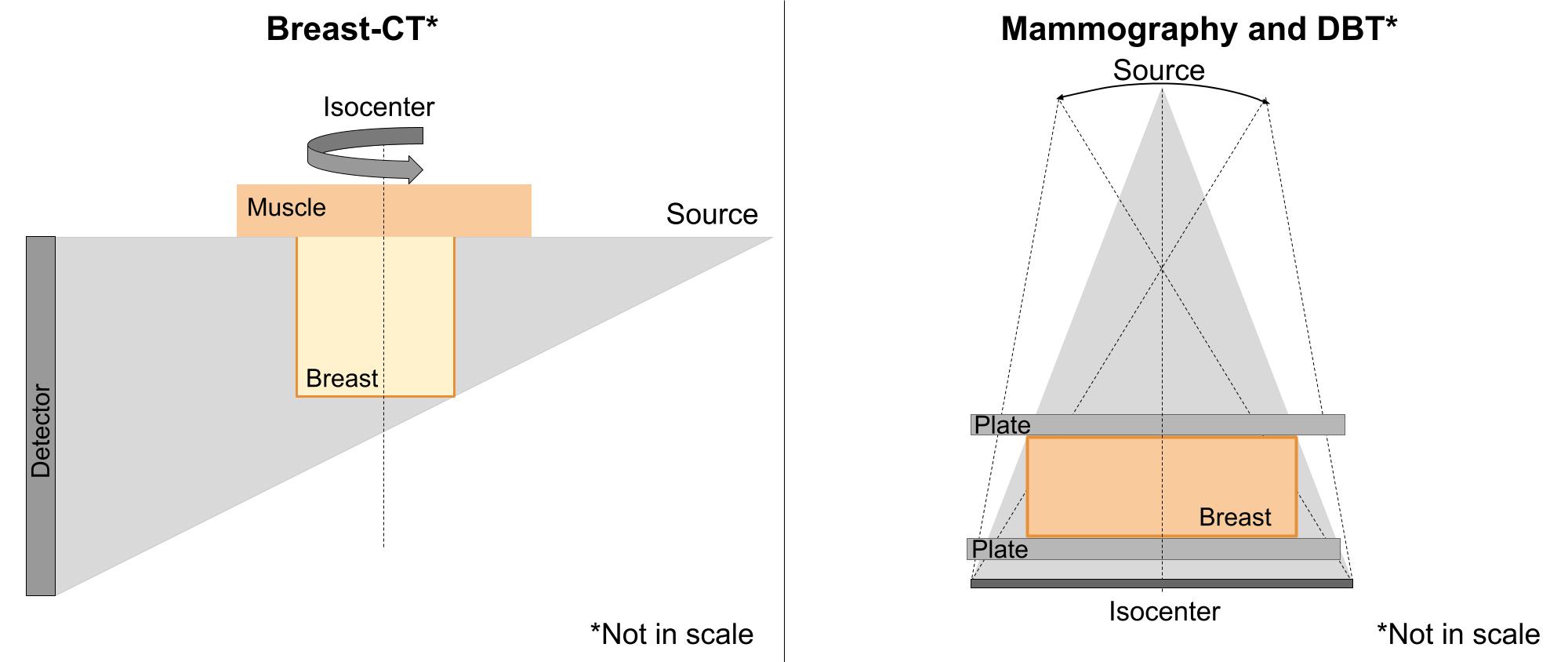}
\caption{Schematic of the geometries used in section \ref{sec:mgd_dgn_val} for the dosimetry validations. Figures not in scale.}
\label{fig:geo_illustration}
\vspace{5mm}
\end{figure}

For the breast-CT validations, the setup was based on the work of Sarno et al.~\cite{Sarno2018}. The breast was modeled as a cylinder with a radius/height of:\ 4~cm/4~cm; 6~cm/9~cm; 9~cm/18~cm, including a 1.5~mm skin layer and the patient chest (a block of muscle tissue, while the original work uses water). For the original work and PENELOPE, the breast was modeled as a homogeneous mixture of adipose-glandular tissues. Meanwhile, for MC-GPU, the geometry consisted of voxelized rectilinear geometries, and the breast model was voxelized with a randomly sampled adipose-glandular distribution. The glandularity varied from 0.1\% to 100\%. The MGD for the heterogeneous model was calculated by summing the energy deposited in glandular voxels divided by their total mass. Meanwhile, the MGD for the homogeneous models was obtained by applying  a  weighting factor (G) \cite{Boone1999,Sarno2018} to the imparted energy in the homogeneous mixture then dividing by the mass of glandular tissue. Afterwards, the breast was replaced by rectangular box of air  (3 $\times$ 1.8 $\times$ 1.1 cm$^{3}$) simulating an ionization chamber (at the isocenter) and the air kerma (\Kair) was scored inside this region. Finally, the Normalized Glandular Dose (\DgNCT) was calculated by the ratio: MGD/\Kair. Therefore, the DgN was compared between the reference work and PENELOPE/MC-GPU results for monoenergetic photons from 10 to 80 keV (5 keV steps).  The total number of  primary photons was on the order of 10$^{8}$. 
Figure \ref{fig:geo_illustration} illustrates the geometry implemented in the simulations for dosimetry validations in this section.

The comparisons with EGSnrc were not included within these tests because the implementation and validation of the code adaptation to perform DBT and breast-CT simulations were beyond the scope of this work.

\subsubsection{Dose distribution comparison for anthropomorphic breast models}
\label{sec:method_mgd}
To quantify the  dose distribution within the breast, a voxelized anthropomorphic breast phantom was generated using the BreastPhantom software \cite{Graff2016} (0.5~mm resolution, 20\% glandularity), and computationally compressed  using the BreastCompress software (with FEBiO \cite{Maas2012}) to 5 cm thick. The breast was irradiated with a W/Rh 28 kV spectrum with a geometry similar to Report-195 (Case III) \cite{Sechopoulos2015} (as described in Table \ref{tab:tomo_bct_mammo_val}, column \textit{Mammography}). Afterwards, the dose in all breast voxels (comprising different materials) was compared between the codes MC-GPU, PENELOPE and EGSnrc to verify the agreement between them, including the dose distribution. The dose was normalized by the number of histories (i.e.\ the number of primary photons that were generated in the source, collimated within the detector field). The total number of  primary photons was on the order of 10$^{10}$. The relative dose difference in a voxel ($\Delta$) was calculated as follows:  
\begin{equation}
\Delta = 100 \times \dfrac{Dg_{ii} - Dg_{i}}{Dg_{i}} \qquad \text{\%},
\label{eq:rel_dif}
\end{equation}
where subscripts refer to MC-GPU  ($i$) and PENELOPE or EGSnrc ($ii$).

\subsection{Phase Space File}

A tracking algorithm was adapted from PENELOPE for MC-GPU (named ``Voxel$\_$intercept'') to identify photons that cross the boundaries of a given voxel from the outside. The ray tracing routine for quadric geometries was implemented to check if the photon intercepts one of the six cube faces. If more than one plane is crossed, the plane closest to the starting point is selected, then the routine stores the partial state variables in memory (energy, position coordinates and direction of movement). After each angular step of the x-ray tube rotation, the information stored in the GPU memory is dumped to disk  in a temporary binary file. When the simulation is finished, a software program (PSFConverter), which was written from an adapted code from penEasy 2019, is called to convert the raw binary file to a format compatible with the IAEA standard \cite{Capote2006}. Data for each particle (position, direction, energy) are stored in 29 bytes.  In order to verify if the framework is set up correctly, three tests described in the following sections were performed.

\subsubsection{Energy distribution comparison}

The simulation of the anthropomorphic phantom (section \ref{sec:method_mgd}) was adapted to  record the energy of photons that entered in a specific voxel near the middle of the breast (simulation description in Table \ref{tab:tomo_bct_mammo_val}). Two spectra were employed (from TASMICS):\ W/Rh 28 kV and W/Cu 49 kV, to represent mammography and CEDM modalities, respectively. The functions to score the energy spectrum of photons were enabled in PENELOPE and EGSnrc. For MC-GPU, the information was retrieved by the generated phase space file. Finally, the photon energy spectra recorded by the three codes were compared.

In addition, the anthropomorphic phantom was downsampled to 2 mm voxels and two phase space files were generated:\ one in MC-GPU and other in PENELOPE for the mammography spectrum. Afterwards, the distribution of the particles' position and direction contained in the phase space files were compared.

\subsubsection{Glandular dose reconstruction}

MC-GPU was used to simulate irradiation of the  anthropomorphic breast phantom in four scenarios:\ (i) mammography; (ii) DBT; (iii) CEDM; (iv) breast-CT (uncompressed breast). For each setup, five phase space files  were recorded in glandular voxels using MC-GPU. Afterwards, the phase space file was loaded in PENELOPE and EGSnrc where the geometry consisted of a single glandular voxel, and it was irradiated in order to score the dose. Therefore, the reconstructed doses from the phase space files in PENELOPE and EGSnrc were compared to the MC-GPU reported doses.

For all modalities, the number of simulated histories in MC-GPU was fine-tuned to yield a mean glandular dose of 4 mGy. For mammography and CEDM, the spectrum was the same as the previous section, while for the breast-CT simulation, the selected spectrum was W/Al 49 kV \cite{Sechopoulos2012}. The number of projections was 120, with a constant number of histories (fixed mAs per scan). For DBT, the selected spectrum was W/Al 31 kV, with 31 projections.

\subsubsection{Practical example}

As an example of  application of the phase space file implementation, a simplified case of multiscale MC simulation was studied and the results of a full simulation performed in PENELOPE was compared to a simulation with MC-GPU plus PENELOPE (using the phase space file approach).

For this, the geometry for the mammography case described in Table \ref{tab:tomo_bct_mammo_val} was implemented. The inner breast tissue was modified to include only adipose tissue (to facilitate the implementation), except in one region at the middle of the breast (a cube of 2~mm sides) where the material was set to glandular tissue. {In} this glandular region, the energy cutoffs for electrons and photons were set to 50~eV to enable a detailed simulation. In addition, the cube {was} sectioned in small sub regions of 10~$\mu$m  side voxels, and the specific energy (energy imparted divided by mass) distribution was scored. In PENELOPE, this simulation was performed in a single step. For MC-GPU, the macroscopic simulation was performed and a phase space file was generated to describe the particles entering in the glandular voxel. Afterwards, the phase space file was loaded in PENELOPE and a detailed simulation was carried out to score the specific energy distribution in the cube subregions, i.e.,   $(10\times10\times10$~$\mu$m$^{3})$. Only subregions more than 50 $\mu$m from the edge of the glandular voxel were considered for the analysis to ensure that electron transport is accurately modeled. A total of 3.2$\times$10$^{11}$ primary photons were simulated.

%--------------------------------------

%-------------------------------------- RESULTS

\section{Results} \label{sec:Results}

\subsection{Digital Breast Tomosynthesis and Breast-CT}

\begin{figure}[htb!]
\includegraphics[width=0.94\textwidth]{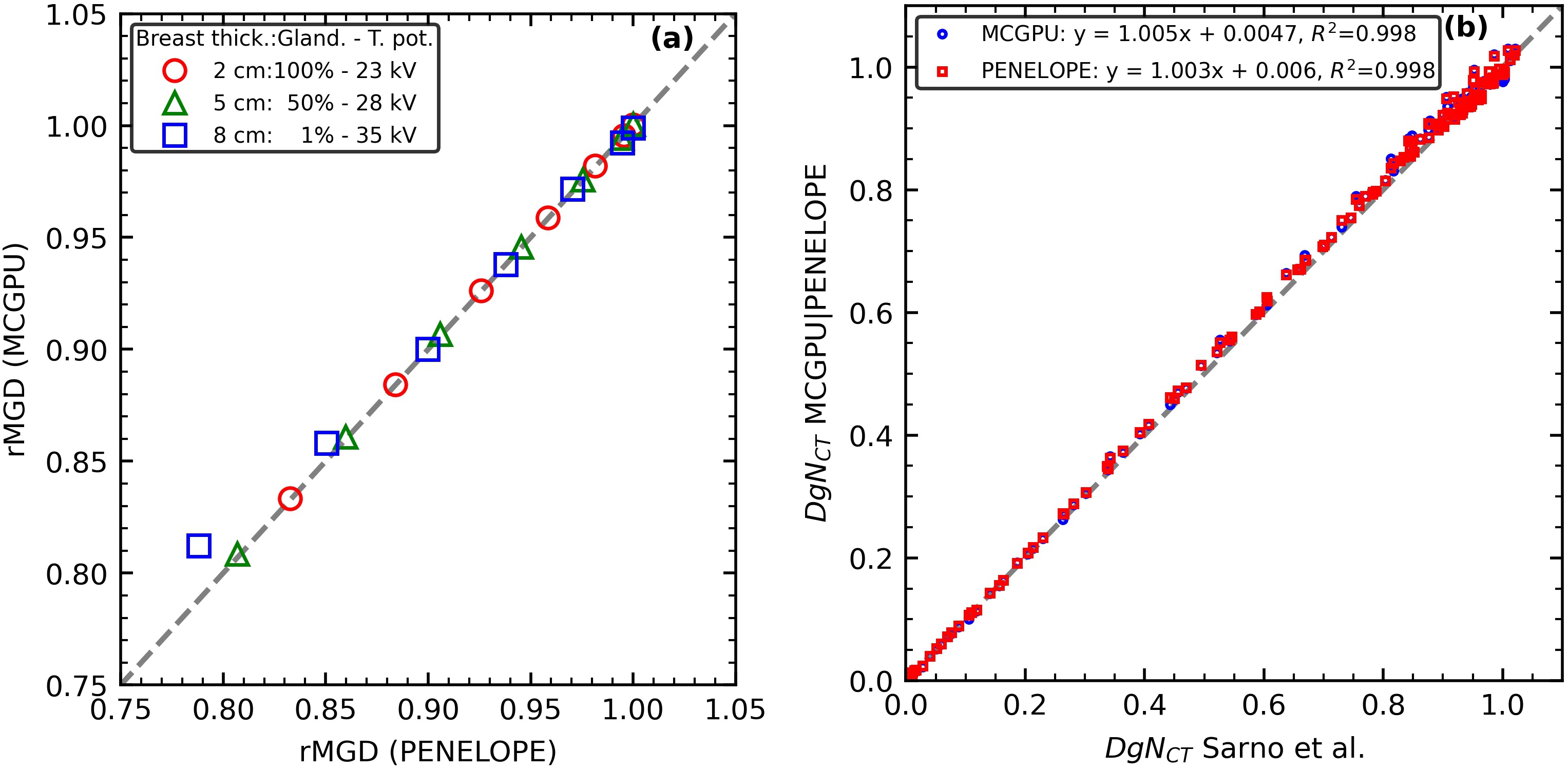}
\caption{(a) Comparison between the relative MGD (rMGD) values for MC-GPU and PENELOPE for DBT for different breast thicknesses, glandularities and tube potentials. The results were normalized by the PENELOPE 0$^{\circ}$ projection MGD value for each breast thickness to obtain the rMGD values. Coefficient of variation: 0.25\% (2$\sigma$). (b) Comparison between {\DgNCT}  values for MC-GPU, PENELOPE and Sarno et al. \cite{Sarno2018} for photon energies between 10 keV and 80 keV. The linear fit quantifies agreement between them, coefficient of variation:\ 0.7\% (2$\sigma$). For both cases, the dashed lines indicate a perfect agreement. }
\label{fig:val_tomo_ct}\vspace{3mm}
\end{figure}

Figure \ref{fig:val_tomo_ct}(a) compares the relative MGD values for the DBT between PENELOPE and MC-GPU for three breast thicknesses with distinct glandular proportions. An excellent agreement was found between the codes with differences smaller than 0.25\% (statistical uncertainties below 0.14\%, 1$\sigma$), except for the 8~cm breast with projection angles 25$^{\circ}$ and 30$^{\circ}$, where the differences were 0.9\% and 3.0\%, respectively. This difference could be explained by the variations on the beam collimation algorithm for the DBT mode among the codes, more specifically, the projected x-ray field fluence at the surface of thicker breasts for high  angles of incidence.  Figure \ref{fig:val_tomo_ct}(b) shows a good agreement between the MC codes and also with the work of Sarno et al. \cite{Sarno2018}, with linear fits close to an ideal line, and an average relative difference of  2.3\%. However, it is important to notice that for low energies (around 10 keV) where the DgN$_{CT}$ is below 0.05, some differences between MC-GPU and Sarno et al.\ were up to 27\%. This could be explained by  the different cross sections used in the codes, the air kerma acquisition geometry and the randomized-sampling of glandular voxels inside the heterogeneous breast phantom. However, those low energies have a negligible impact in the dose when integrating over a breast-CT spectrum. For PENELOPE and MC-GPU, the average and maximum DgN$_{CT}$  relative  differences were 0.87\% and 12.6\%, respectively.

Regarding performance, for illustration, MC-GPU and PENELOPE (Ryzen 1700X, using only 1 core) presented a simulation speed of 1.76$\times$10$^{7}$ and 1.44$\times$10$^{5}$ histories/s, respectively for a breast-CT simulation of 50 keV monoenergetic photons and a large breast (50\% glandular tissue). The ratio between the simulation speed achieved  for MC-GPU and PENELOPE codes goes from approximately 242 at 10 keV down to 121 at 80 keV.

\subsection{Dose distribution}

\begin{figure}[htbp]
\includegraphics[width=0.94\textwidth]{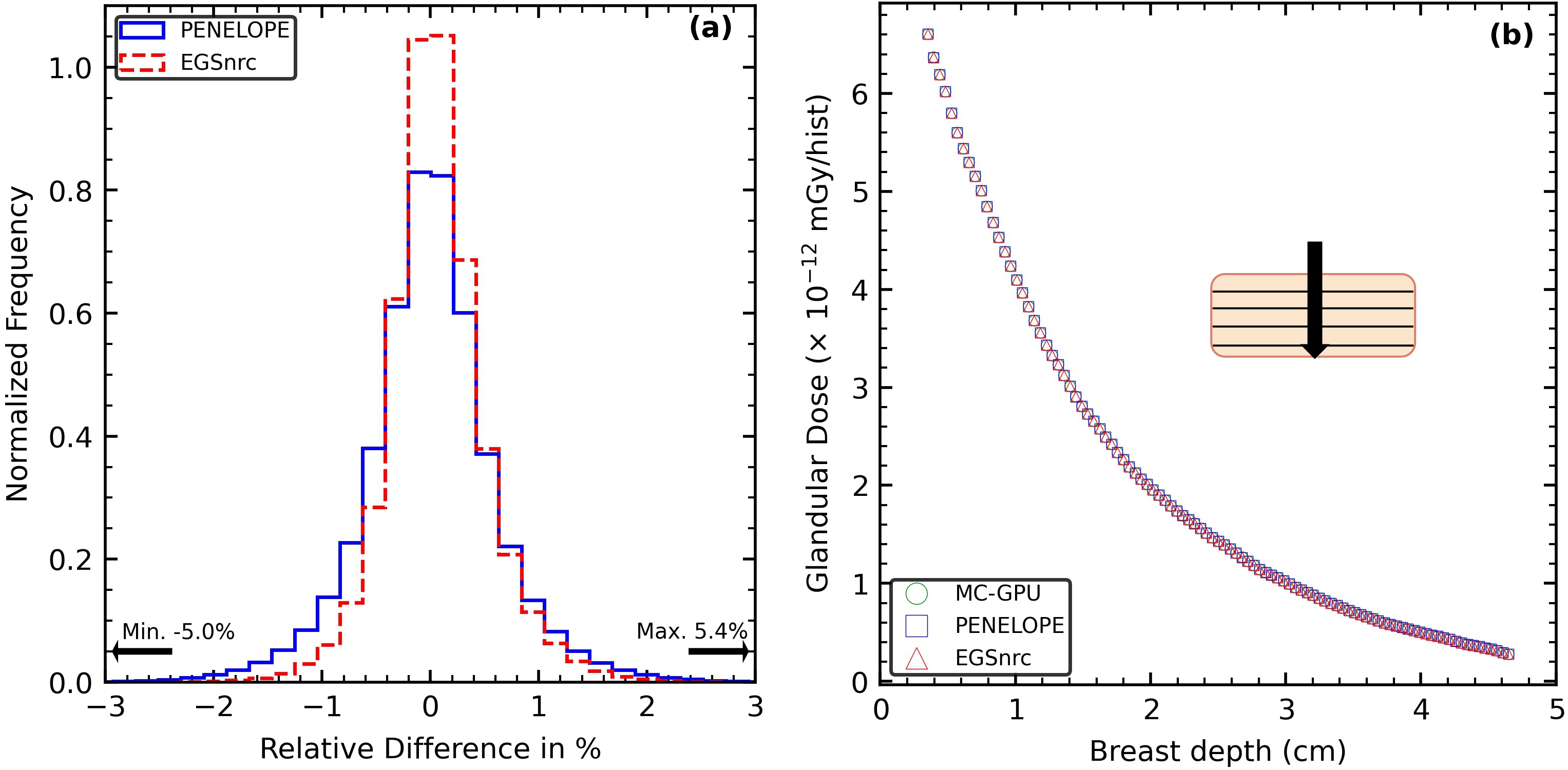}
\caption{(a) Relative difference between MC-GPU and PENELOPE/EGSnrc for the breast dose voxels.  The arrows indicate the minimum and maximum values. Maximum coefficient of variation:\ 2\%. (b) Glandular dose as function of the breast depth for different MC codes. Each point represents the average value for all glandular voxels in a particular depth. The reference plane is exemplified by the insert. Imaging modality:\ Mammography.}
\label{fig:compare_glanddose}
\vspace{2mm}
\end{figure}

The relative difference between the breast dose distribution in voxels for PENELOPE and EGSnrc compared to MC-GPU are shown in Figure \ref{fig:compare_glanddose}(a). The differences resemble a normal distribution, without an apparent offset (i.e.\ centered near zero), which is consistent with the statistical uncertainty of the values. The uncertainty obtained with PENELOPE were higher compared to the other codes  due to the longer computation times (smaller number of available processors). Nevertheless, the majority of the differences are contained within the -1 to 1\% interval. The glandular dose as function of the breast depth is shown in Figure \ref{fig:compare_glanddose}(b). An excellent agreement was found between the codes, with differences smaller than 0.35\%. 
The voxel with maximum dose (excluding air and the plates) was the same for all three codes, which is located at the top of the breast, with differences lower than 0.4\% between the dose values.

\subsection{Phase Space File:\ Photon energy spectrum}

\begin{figure}[htpb]
\includegraphics[width=0.94\textwidth]{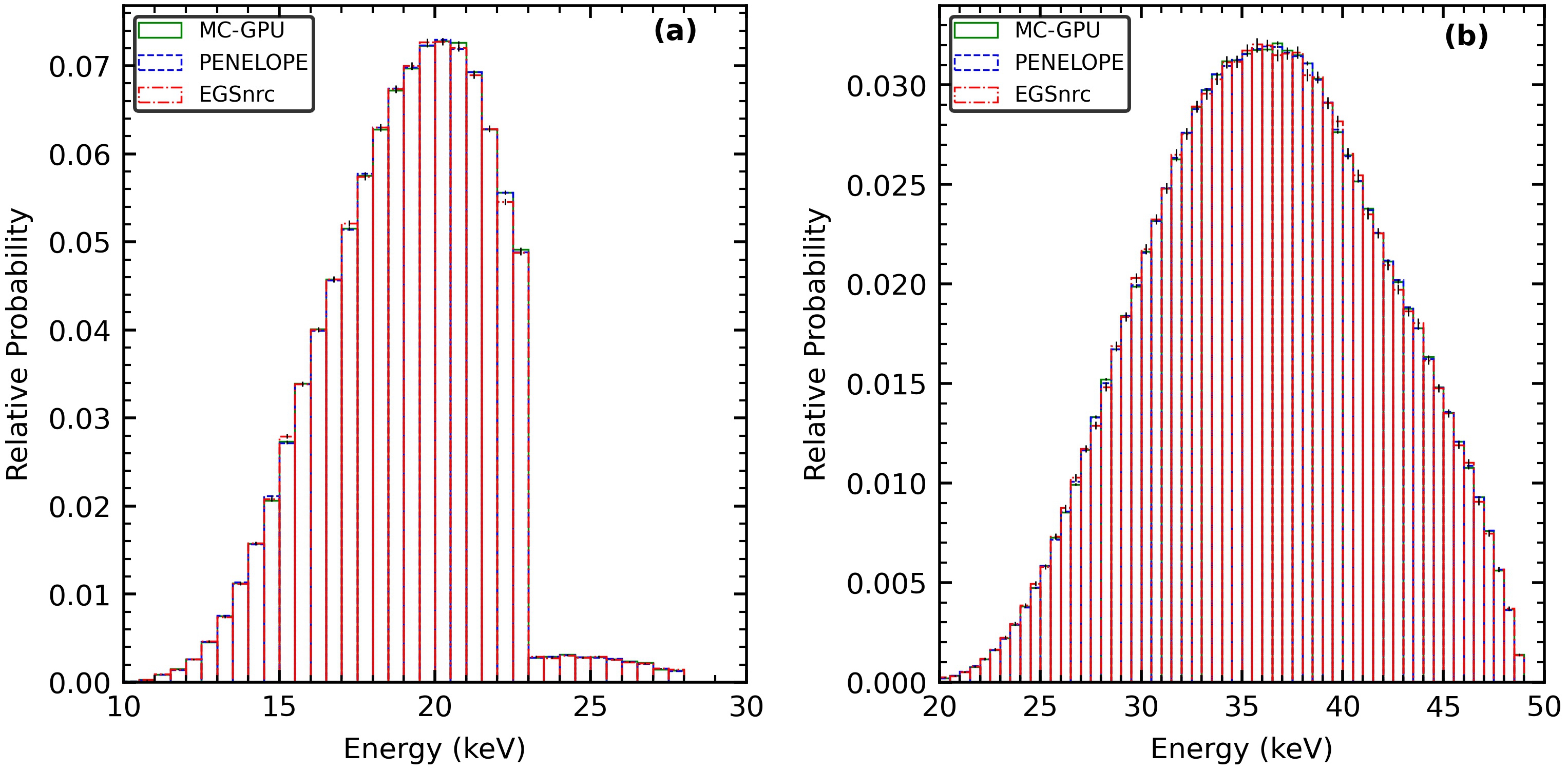}
\caption{Spectra of photons entering in a glandular voxel inside the breast recorded in different MC codes. The simulations were performed with the following spectra:\ (a) W/Rh 28 kV and (b) W/Cu 49 kV.}
\label{fig:compare_spectrum} \vspace{2mm}
\end{figure}

In order to validate the algorithm implemented in MC-GPU  to generate the phase space files, the recorded spectrum of photons entering in a voxel was compared with the PENELOPE and EGSnrc MC codes. The results are shown in Figure \ref{fig:compare_spectrum} where it can be observed that the relative probability is similar between the codes within the estimated statistical uncertainty for both x-ray spectra. 
The bins below 10 keV were omitted due to the relative low probability and, consequently, the low impact in the results. The average relative differences for W/Rh 28 kV (10 keV threshold) and W/Cu 49 kV (20 keV threshold) between MC-GPU and PENELOPE/EGSnrc  were lower than 2.5\%.

\subsection{Phase Space File:\ Glandular Dose reconstruction}

The  glandular dose values obtained in MC-GPU (full simulation)  compared to those obtained within PENELOPE and EGSnrc (phase space files) are shown in Table \ref{tab:psf_dose} with their respective statistical uncertainties. The average $\Delta$ between MC-GPU and PENELOPE/EGSnrc was 0.22\%/0.18\%, with maximum $\Delta$ values of 0.63\%/0.43\%. Considering that $1\sigma$ statistical uncertainty is approximately 0.3\% for both PENELOPE and EGSnrc, and 0.004\% for MC-GPU, the glandular dose values computed using phase space files (PENELOPE, EGSnrc) are in good agreement with those from full simulation (MC-GPU).

\begin{table}[t!]
\caption{Dose in five distinct glandular voxels (ROI) obtained directly with MC-GPU, and reconstructed from phase space files for EGSnrc and PENELOPE. Values in mGy. The statistical uncertainty (in mGy) is indicated by the values in parentheses.}
\centering
\label{tab:psf_dose}
\begin{tabular}{cccc}
\hline
\hline
\multicolumn{1}{l}{ROI} & MC-GPU    & PENELOPE & EGSnrc   \\
\hline
\multicolumn{1}{l}{}  & \multicolumn{3}{c}{Mammography} \\ 

1 & 3.4230(1) & 3.424(9) & 3.416(9)  \\
2 & 3.7656(1) & 3.769(9) & 3.760(9)  \\
3 & 3.3138(1) & 3.302(9) & 3.296(9) \\
4 & 3.3027(1) & 3.295(9) & 3.280(9) \\
5 & 3.2874(1) & 3.299(9) & 3.296(9)  \\
\\
\multicolumn{1}{l}{}  & \multicolumn{3}{c}{DBT}         \\
1 &  3.4508(1) & 3.442(9) & 3.442(9) \\
2 &  3.6832(1) & 3.661(9) & 3.661(9) \\
3 &  3.3716(1) & 3.362(9) & 3.362(9) \\
4 &  3.4411(1) & 3.433(9) & 3.433(9) \\
5 &  3.4237(1) & 3.409(9) & 3.409(9) \\
\\
\multicolumn{1}{l}{}    & \multicolumn{3}{c}{CEDM}        \\
1 & 4.4531(2) & 4.46(1)  & 4.46(1)    \\
2 & 4.1154(1) & 4.13(1)  & 4.13(1)    \\
3 & 4.1303(1) & 4.15(1)  & 4.15(1)    \\
4 & 4.2368(1) & 4.26(1)  & 4.25(1)   \\
5  & 4.3637(2) & 4.36(1)  & 4.36(1)  \\
\\
\multicolumn{1}{l}{}    &\multicolumn{3}{c}{Breast-CT}   \\
1 &  3.4240(1) & 3.44(1)  & 3.43(1)  \\
2 &  3.7656(1) & 3.78(1)  & 3.76(1)  \\
3  & 3.3138(1) & 3.311(8) & 3.31(1)  \\
4  & 3.3027(1) & 3.30(1)  & 3.32(1)  \\
5   & 3.2874(1) & 3.308(9) & 3.29(1) \\
\hline
\hline
\end{tabular}
\end{table}

\subsection{Practical example}
\label{sec:practical_examples_res}

\begin{figure}[htpb]
\includegraphics[width=0.95\textwidth]{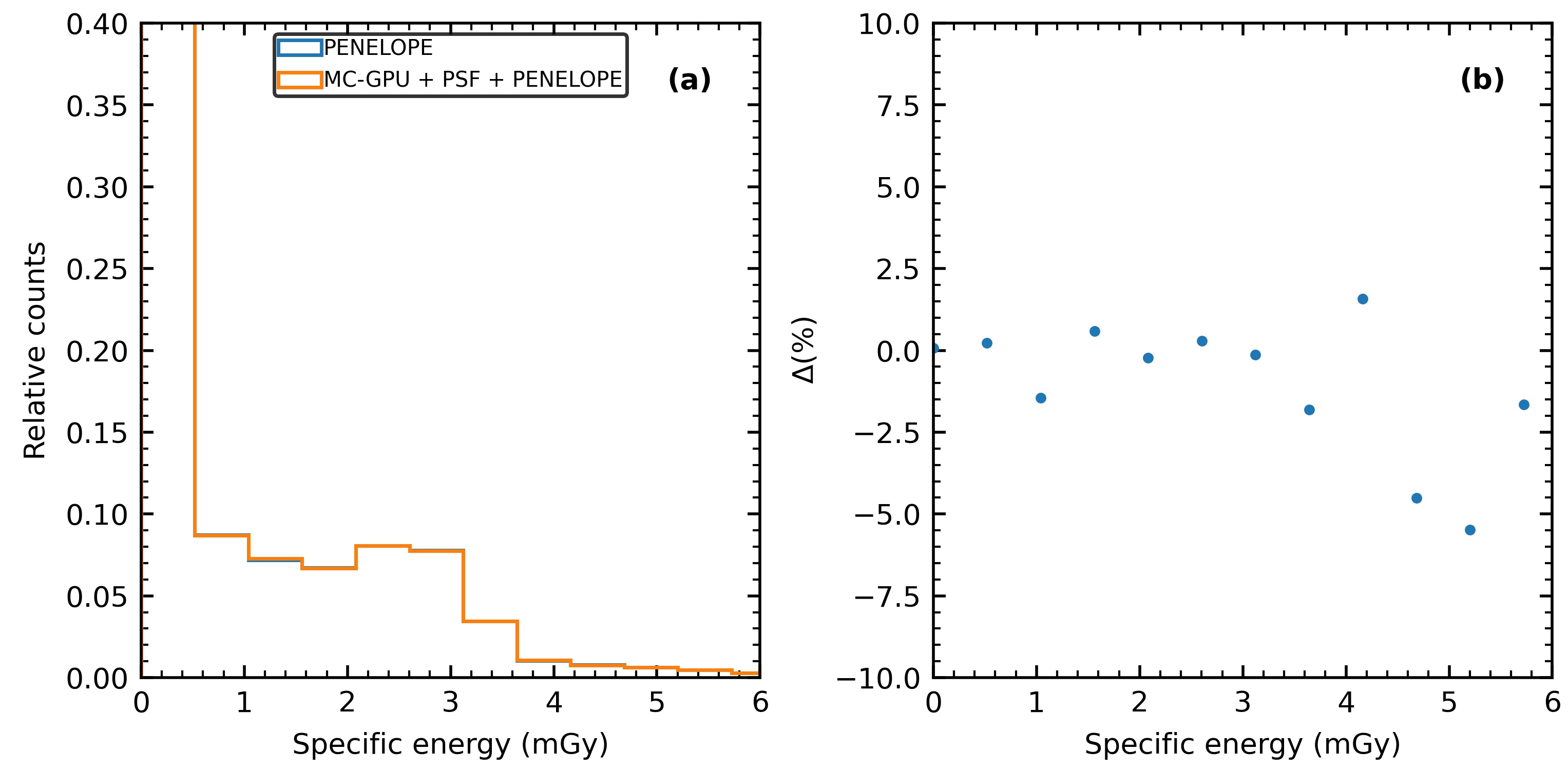}
 \caption{(a) Specific energy distribution obtained in a full simulation within PENELOPE and using the phase space file approach (MC-GPU + PSF + PENELOPE). (b) Relative differences ($\Delta$)  between both approaches for each bin. Values of specific energy higher than 6 mGy where excluded due to their low probabilities.}
\label{fig:psf_full_simulation} \vspace{3mm}
\end{figure}

Figure \ref{fig:psf_full_simulation} compares the results from the multiscale simulation using PENELOPE and the phase space file using MC-GPU plus PENELOPE method proposed in this work, showing an excellent agreement. The PENELOPE simulation took approximately 65.4 hours (Ryzen 1700X, using 8 cores) to finish. Meanwhile, the whole process of MC-GPU generating the phase space file then simulating in PENELOPE took approximately 3.2 hours, a speed-up of approximately 20 times. The time spent in file manipulations was on the order of seconds, 6\% of the time in the PENELOPE simulation and the rest in MC-GPU.  It is important to notice that by turning on the additional phase space files calculations in MC-GPU, the performance was slowed by approximately 30\%. The generated phase space file size was approximately 450 megabytes, which resulted in a negligible impact on the performance of the calculations ($\approx$ hours) due to disk read/write operations ($\approx$ seconds).

%-----------------------------------

%----------------------------------- Discussion
\section{Discussion} \label{sec:Discussion}

MC-GPU simulates photon transport through matter with physics models based on PENELOPE \cite{Badal2009}, with minor modifications. As seen in the results, there is a good overall agreement between MC-GPU and the other considered MC codes. Moreover, a comparison with EGSnrc and the work of Sarno et al.\ \cite{Sarno2018} (which used a program based on GEANT4) was also included. In the latter case, high discrepancies were observed at very low energies, but could be explained by the differences in the cross sections of the codes and minor modifications in the geometry. In breast imaging simulations, it is often assumed that the electrons are locally deposited \cite{Sechopoulos2015, Sarno2018}  due to mm-to-cm length scales of simulated objects, significantly larger than the short range of electrons at low energies  (from 0.05~$\mu$m  at 1~keV to 144~$\mu$m at 100~keV, CSDA in liquid water).  Thus, MC-GPU may be used for efficient dosimetric simulations, enabling a large number of simulations with limited compute cluster resources.  A generalized  comparison of simulation speeds shows that using MC-GPU in a GeForce GTX 1060 (NVIDIA, USA) had a performance 40 times greater than PENELOPE in a Core i7 7700 (Intel, USA) processor (using all cores). Although limited (since we are comparing CPU to GPU), these results at least show the performance improvements that could be achieved when desktops (with a limited number of CPUs) are used in MC simulations. The need to optimize simulation efficiency becomes important especially with recent studies focusing in complex breast models \cite{Hernandez2019, Badano2018, Mettivier2020, Sechopoulos2012, Hernandez2015}, where a high computation power is needed. The MC-GPU code also supports the use of a search tree in the tracking algorithm which greatly reduces the amount of memory to store high-resolution phantoms. This is highly efficient compared to the parallelism implemented in some MC simulations where the jobs do not  share memory and the same breast phantom must be loaded for every job.

The phase space file functionality implemented in MC-GPU was developed to support future multiscale studies of breast dosimetry \cite{Oliver2019}, where the macroscopic scale would be simulated in MC-GPU and the microscopic scale  in a different code with electron transport, such as PENELOPE or EGSnrc. 
The effects of potentially missed secondary photons, in this application, is small\cite{Sarno2017b} and can be disregarded since low energy photons are used for x-ray imaging of the breast, associated with low atomic number of breast tissues.

Moreover, this routine could be adapted for other purposes, such as simulating energy deposition in a detector. There is a trade off between the simulation speed and the number of particles being scored. The user should optimize between the size of the scoring region and the total number of particles entering the region for efficiency and in order to limit the size of the phase space file considering the overhead time to write and load them. It is important to notice that MC-GPU does not simulate electrons and does not have routines to calculate fluorescence effects in the materials. Since the effective atomic number of breast tissues is relatively low, the probability of fluorescence is negligible. However, this might be needed in other applications. Another interesting feature implemented in the phase space file generating algorithm is the option to not kill  particles that enter the volume of interest. This is particularly useful for the application discussed in this work of recreating a dose in a voxel because it ensures that backscatter photons are included.

Preliminary tests (not included in this work) show that for a mammography simulation and a voxel in the middle of the breast, 2\% and 0.5\% of photons are missed if photons are killed when they enter the volume of interest
for voxels with 2 mm and 0.5 mm side length, respectively.

For the practical example (Section \ref{sec:practical_examples_res}), the average dose in adipose tissue was 0.54~mGy, almost on the same order of magnitude found in real mammography imaging. Thus, it is expected that the size of the phase space files in this type of multiscale studies would be in the worst case scenario of a few gigabytes, which is still viable with most current hardware available.

%-----------------------------------

%----------------------------------- Conclusion

\section{Conclusion} \label{sec:Conclusion}

Recent studies of breast dosimetry employ complex breast models with realistic features, presenting considerable demands on computing power. The present article demonstrates that MC-GPU is suitable for carrying out accurate MC dosimetric evaluations for different x-ray breast imaging modalities. Moreover, the option to record phase space files in specific regions of the geometry has been successfully implemented.  This development will enable future studies of energy deposition on different scales by also employing an MC code that models coupled electron-photon transport, e.g., the relation between dose in macroscopic models and the specific energy imparted in cells.   In theory, any code that allows the IAEA phase space file format is compatible to work with the files generated with MC-GPU. The authors will release the modified MC-GPU code in a digital repository (https://github.com/rtmass/MCGPU-PSF). Future studies could expand these applications to other x-ray imaging techniques besides the breast and other low x-ray energies applications.

%-----------------------------------

\section*{Conflicts of interest}
The authors have no conflicts of interest.

% -----------------------------------------------------
\section*{Acknowledgments}

This work was supported by  Fundação de Amparo à Pesquisa do Estado de São Paulo (FAPESP) - Project numbers 2015/21873-8, 2016/15366-9 and {2018/05982-0}, Ministério da Ciência, Tecnologia e Inovações e Conselho Nacional de Desenvolvimento Científico e Tecnológico (CNPq) - Project number 140155/2019-8, Coordenação de Aperfeiçoamento de Pessoal de Nível Superior Brasil (CAPES) - Finance Code 001, AAPM International Training and Research Coordination scholarship, Emerging Leaders in America Program scholarship with support of the Government of Canada, the Canada Research Chairs program, and the Natural Sciences and Engineering Research Council of Canada. The authors would like to thank Iymad Mansour for sharing the code to score photon spectrum in a voxel for egs$\_$brachy.  

\section*{Data availability}

The modifications implemented in MC-GPU for this study are available in a digital repository (https://github.com/rtmass/MCGPU-PSF). Specific data are available on reasonable request from the correspondence author.

%%%%%%%%%%%%%%%%%%%%%%%%%%%%%%%%%%%%%%%%%%%%%%%%%%%%%%%%%%%%%%%%%%%%%%%%%%%%%%%%%%%%%%%%%%%%%
% BIBLIOGRAPHY
%%%%%%%%%%%%%%%%%%%%%%%%%%%%%%%%%%%%%%%%%%%%%%%%%%%%%%%%%%%%%%%%%%%%%%%%%%%%%%%%%%%%%%%%%%%%%
% Good for drafts
\setlength{\baselineskip}{0.55cm}
% File path to .bib containing references
\bibliographystyle{medphy}
\bibliography{references}

\begin{thebibliography}{10}

\bibitem{Dance1990}
D.~R. Dance,
\newblock {Monte Carlo calculation of conversion factors for the estimation of
  mean glandular breast dose},
\newblock Physics in Medicine and Biology {\bf 35}, 1211--9 (1990).

\bibitem{Wu1991}
X.~Wu, G.~T. Barnes, and D.~M. Tucker,
\newblock {Spectral dependence of glandular tissue dose in screen-film
  mammography},
\newblock Radiology {\bf 179}, 143--148 (1991).

\bibitem{Boone1999}
J.~M. Boone,
\newblock {Glandular breast dose for monoenergetic and high-energy x-ray beams:
  Monte Carlo assessment},
\newblock Radiology {\bf 213}, 23--37 (1999).

\bibitem{Sarno2018}
A.~Sarno, G.~Mettivier, R.~M. Tucciariello, K.~Bliznakova, J.~M. Boone,
  I.~Sechopoulos, F.~{Di Lillo}, and P.~Russo,
\newblock {Monte Carlo evaluation of glandular dose in cone-beam X-ray computed
  tomography dedicated to the breast:\ Homogeneous and heterogeneous breast
  models},
\newblock Physica Medica {\bf 51}, 99--107 (2018).

\bibitem{Dance2016}
D.~R. Dance and I.~Sechopoulos,
\newblock {Dosimetry in x-ray-based breast imaging},
\newblock Physics in Medicine and Biology {\bf 61}, R271--R304 (2016).

\bibitem{Mettivier2020}
F.~di~Franco, A.~Sarno, G.~Mettivier, A.~M. Hernandez, K.~Bliznakova, J.~M.
  Boone, and P.~Russo,
\newblock {GEANT4 Monte Carlo simulations for virtual clinical trials in breast
  X-ray imaging:\ Proof of concept},
\newblock Physica Medica {\bf 74}, 133--142 (2020).

\bibitem{Hernandez2015}
A.~M. Hernandez, J.~A. Seibert, and J.~M. Boone,
\newblock {Breast dose in mammography is about 30{\%} lower when realistic
  heterogeneous glandular distributions are considered},
\newblock Medical Physics {\bf 42}, 6337--6348 (2015).

\bibitem{Hernandez2019}
A.~M. Hernandez, A.~E. Becker, and J.~M. Boone,
\newblock {Updated breast CT dose coefficients (DgN CT) using patient-derived
  breast shapes and heterogeneous fibroglandular distributions},
\newblock Medical Physics {\bf 46}, 1455--1466 (2019).

\bibitem{Badal2009}
A.~Badal and A.~Badano,
\newblock {Accelerating Monte Carlo simulations of photon transport in a
  voxelized geometry using a massively parallel graphics processing unit},
\newblock Medical Physics {\bf 36}, 4878--4880 (2009).

\bibitem{Badano2018}
A.~Badano, C.~G. Graff, A.~Badal, D.~Sharma, R.~Zeng, F.~W. Samuelson, S.~J.
  Glick, and K.~J. Myers,
\newblock {Evaluation of Digital Breast Tomosynthesis as replacement of
  full-field digital mammography using an in silico imaging trial},
\newblock JAMA Network Open {\bf 1}, e185474 (2018).

\bibitem{Badal2020}
A.~Badal, D.~Sharma, C.~G. Graff, R.~Zeng, and A.~Badano,
\newblock {Mammography and breast tomosynthesis simulator for virtual clinical
  trials},
\newblock Computer Physics Communications {\bf 261}, 107779 (2020).

\bibitem{Ghammraoui2014}
B.~Ghammraoui and A.~Badal,
\newblock {Monte Carlo simulation of novel breast imaging modalities based on
  coherent x-ray scattering},
\newblock Physics in Medicine and Biology {\bf 59}, 3501--3516 (2014).

\bibitem{Ghammraoui2014b}
B.~Ghammraoui, R.~Peng, I.~Suarez, C.~Bettolo, and A.~Badal,
\newblock {Including the effect of molecular interference in the coherent x-ray
  scattering modeling in MC-GPU and PENELOPE for the study of novel breast
  imaging modalities},
\newblock in {\em Medical Imaging 2014: Physics of Medical Imaging}, edited by
  B.~R. Whiting and C.~Hoeschen, volume 9033, page 90334N, SPIE, 2014.

\bibitem{FernandezBosman2021}
D.~{Fern{\'{a}}ndez Bosman}, V.~{Garc{\'{i}}a Balcaza}, C.~Delgado,
  S.~Principi, M.~A. Duch, and M.~Ginjaume,
\newblock {Validation of the MC-GPU Monte Carlo code against the
  PENELOPE/penEasy code system and benchmarking against experimental conditions
  for typical radiation qualities and setups in interventional radiology and
  cardiology},
\newblock Physica Medica {\bf 82}, 64--71 (2021).

\bibitem{Garcia2021}
V.~García~Balcaza, A.~Camp, A.~Badal, M.~Andersson, A.~Almen, M.~Ginjaume, and
  M.~Duch,
\newblock Fast Monte Carlo codes for occupational dosimetry in interventional
  radiology,
\newblock Physica Medica {\bf 85}, 166--174.

\bibitem{Sharma2019b}
S.~Sharma, A.~Kapadia, W.~Fu, E.~Abadi, W.~P. Segars, and E.~Samei,
\newblock A real-time Monte Carlo tool for individualized dose estimations in
  clinical {CT},
\newblock Physics in Medicine \& Biology {\bf 64}, 215020 (2019).

\bibitem{Bert2013}
J.~Bert, H.~Perez-Ponce, Z.~E. Bitar, S.~Jan, Y.~Boursier, D.~Vintache,
  A.~Bonissent, C.~Morel, D.~Brasse, and D.~Visvikis,
\newblock Geant4-based {Monte} {Carlo} simulations on {GPU} for medical
  applications,
\newblock Physics in Medicine and Biology {\bf 58}, 5593--5611 (2013),
\newblock Publisher: IOP Publishing.

\bibitem{Lippuner2011}
J.~Lippuner and I.~A. Elbakri,
\newblock A {GPU} implementation of {EGSnrc}'s {Monte} {Carlo} photon transport
  for imaging applications,
\newblock Physics in Medicine and Biology {\bf 56}, 7145--7162 (2011).

\bibitem{Tsai2020}
M.-Y. Tsai, Z.~Tian, N.~Qin, C.~Yan, Y.~Lai, S.-H. Hung, Y.~Chi, and X.~Jia,
\newblock A new open-source {GPU}-based microscopic {Monte} {Carlo} simulation
  tool for the calculations of {DNA} damages caused by ionizing radiation -
  {Part} {I}:\ {Core} algorithm and validation,
\newblock Medical Physics {\bf 47}, 1958--1970 (2020).

\bibitem{Lai2020}
Y.~Lai, M.-Y. Tsai, Z.~Tian, N.~Qin, C.~Yan, S.-H. Hung, Y.~Chi, and X.~Jia,
\newblock A new open-source {GPU}-based microscopic {Monte} {Carlo} simulation
  tool for the calculations of {DNA} damages caused by ionizing radiation -
  {Part} {II}:\ sensitivity and uncertainty analysis,
\newblock Medical Physics {\bf 47}, 1971--1982 (2020).

\bibitem{Sechopoulos2015}
I.~Sechopoulos, E.~S.~M. Ali, A.~Badal, A.~Badano, J.~M. Boone, I.~S.
  Kyprianou, E.~Mainegra-Hing, K.~L. McMillan, M.~F. McNitt-Gray, D.~W.~O.
  Rogers, E.~Samei, and A.~C. Turner,
\newblock {Monte Carlo reference data sets for imaging research: Executive
  summary of the report of AAPM Research Committee Task Group 195},
\newblock Medical Physics {\bf 42}, 5679--5691 (2015).

\bibitem{Sarno2017b}
A.~Sarno, G.~Mettivier, F.~{Di Lillo}, and P.~Russo,
\newblock {A Monte Carlo study of monoenergetic and polyenergetic normalized
  glandular dose (DgN) coefficients in mammography},
\newblock Physics in Medicine and Biology {\bf 62}, 306--325 (2017).

\bibitem{Oliver2019}
P.~A.~K. Oliver and R.~M. Thomson,
\newblock Investigating energy deposition in glandular tissues for mammography
  using multiscale Monte Carlo simulations,
\newblock Medical Physics {\bf 46}, 1426--1436 (2019).

\bibitem{Badal2021}
A.~Badal, D.~Sharma, C.~G. Graff, R.~Zeng, and A.~Badano,
\newblock {Mammography and breast tomosynthesis simulator for virtual clinical
  trials},
\newblock Computer Physics Communications {\bf 261}, 107779 (2021).

\bibitem{Salvat2019}
F.~Salvat,
\newblock {PENELOPE-2018: A code system for Monte Carlo simulations of electron
  and photon transport}, 2019.

\bibitem{Massera2018}
R.~Massera and A.~Tomal,
\newblock {Skin models and their impact on mean glandular dose in mammography},
\newblock Physica Medica {\bf 51}, 38--47 (2018).

\bibitem{Sempau2011}
J.~Sempau, A.~Badal, and L.~Brualla,
\newblock {A PENELOPE -based system for the automated Monte Carlo simulation of
  clinacs and voxelized geometries-application to far-from-axis fields},
\newblock Medical Physics {\bf 38}, 5887--5895 (2011).

\bibitem{Chamberland2016}
M.~J.~P. Chamberland, R.~E.~P. Taylor, D.~W.~O. Rogers, and R.~M. Thomson,
\newblock egs{\_}brachy:\ a versatile and fast Monte Carlo code for
  brachytherapy,
\newblock Physics in Medicine and Biology {\bf 61}, 8214--8231 (2016).

\bibitem{Sempau2001}
J.~Sempau, A.~Sánchez-Reyes, F.~Salvat, H.~O.~b. Tahar, S.~B. Jiang, and J.~M.
  Fernández-Varea,
\newblock Monte {Carlo} simulation of electron beams from an accelerator head
  using {PENELOPE},
\newblock Physics in Medicine and Biology {\bf 46}, 1163--1186 (2001).

\bibitem{Walters2002}
B.~R.~B. Walters, I.~Kawrakow, and D.~W.~O. Rogers,
\newblock History by history statistical estimators in the {BEAM} code system,
\newblock Medical Physics {\bf 29}, 2745--2752 (2002).

\bibitem{Hammerstein1979}
R.~G. Hammerstein, D.~W. Miller, D.~R. White, M.~E. Masterson, H.~Q. Woodard,
  and J.~S. Laughlin,
\newblock {Absorbed radiation dose in mammography},
\newblock Radiology {\bf 130}, 485--491 (1979).

\bibitem{Woodard1986}
H.~Q. Woodard and D.~R. White,
\newblock {The composition of body tissues},
\newblock The British Journal of Radiology {\bf 59}, 1209--1218 (1986).

\bibitem{Hubbell2004}
M.~Berger, J.~Hubbell, S.~Seltzer, J.~Chang, J.~Coursey, R.~Sukumar, D.~Zucker,
  and K.~Olsen,
\newblock {XCOM: Photon Cross Section Database. National Institute of Standards
  and Technology (NIST)},
\newblock (2010).

\bibitem{Sechopoulos2014}
I.~Sechopoulos, J.~M. Sabol, J.~Berglund, W.~E. Bolch, L.~Brateman,
  E.~Christodoulou, M.~Flynn, W.~Geiser, M.~Goodsitt, A.~{Kyle Jones}, J.~Y.
  Lo, A.~D.~A. Maidment, K.~Nishino, A.~Nosratieh, B.~Ren, W.~{Paul Segars},
  and M.~{Von Tiedemann},
\newblock {Radiation dosimetry in digital breast tomosynthesis: Report of AAPM
  Tomosynthesis Subcommittee Task Group 223},
\newblock Medical Physics {\bf 41}, 091501 (2014).

\bibitem{Hernandez2017}
A.~M. Hernandez, J.~A. Seibert, A.~Nosratieh, and J.~M. Boone,
\newblock {Generation and analysis of clinically relevant breast imaging x-ray
  spectra},
\newblock Medical Physics {\bf 44}, 2148--2160 (2017).

\bibitem{Graff2016}
C.~G. Graff,
\newblock {A new, open-source, multi-modality digital breast phantom},
\newblock in {\em Medical Imaging 2016: Physics of Medical Imaging}, edited by
  D.~Kontos, T.~G. Flohr, and J.~Y. Lo, volume 9783, page 978309, SPIE, 2016.

\bibitem{Maas2012}
S.~A. Maas, B.~J. Ellis, G.~A. Ateshian, and J.~A. Weiss,
\newblock {FEBio:\ Finite elements for biomechanics},
\newblock Journal of Biomechanical Engineering {\bf 134} (2012).

\bibitem{Capote2006}
R.~Capote, R.~Jeraj, C.~M. Ma, D.~W.~O. Rogers, F.~Sanchez-Doblado, J.~Sempau,
  J.~Seuntjens, and J.~V. Siebers,
\newblock {Phase-space database for external beam radiotherapy Summary report
  of a consultants' meeting},
\newblock Technical report, International Atomic Energy Agency (IAEA), 2006.

\bibitem{Sechopoulos2012}
I.~Sechopoulos, K.~Bliznakova, X.~Qin, B.~Fei, and S.~S.~J. Feng,
\newblock {Characterization of the homogeneous tissue mixture approximation in
  breast imaging dosimetry},
\newblock Medical Physics {\bf 39}, 5050--5059 (2012).

\end{thebibliography}
%%\section{Bibliography}
%
%% File path to .bst containing biblio style

\end{document}